\documentclass[aps, pre, superscriptaddress, numerical, amsmath, amssymb, twocolumn, preprintnumbers, longbibliography
]{revtex4-1}

\usepackage{graphicx}% Include figure files
\usepackage[usenames]{color}
\usepackage{mathrsfs}
\usepackage{dcolumn}% Align table columns on decimal point
\usepackage{bm}

\usepackage{amsthm}
\theoremstyle{definition}
\newtheorem{remark}{Remark}

\newcommand{\ri}{\mathrm{i}}

\DeclareMathOperator{\Imag}{Im}
\DeclareMathOperator{\Real}{Re}

\usepackage[colorlinks=true,linkcolor=blue,breaklinks=true]{hyperref}

\begin{document}

%\linenumbers % Commence numbering lines

%\preprint{\textit{To be submitted to: Phys.\ Rev.\  Appl.}}

\title{Comments on ``Scattering Cancellation-Based Cloaking for the Maxwell--Cattaneo Heat Waves''}% Force line breaks with \\
%\thanks{Footnote to title of article.}

\author{Ivan C.\ Christov}%\thanks{Author to whom correspondence should be addressed.}
%Electronic mail 
\email{christov@purdue.edu}
\homepage{http://christov.tmnt-lab.org}
\affiliation{School of Mechanical Engineering, Purdue University, West Lafayette, Indiana 47907, USA}

\date{\today}% It is always \today, today,
             % but any date may be explicitly specified

\begin{abstract}
A number of errors, both mathematical and  conceptual, are identified, in a recent article by Farhat \textit{et al.}\  [Phys.\ Rev.\ Appl.\ \textbf{11}, 044089 (2019)] on cloaking of thermal waves in solids, and corrected.  The differences between the two thermal flux laws considered in the latter article are also critically discussed, specifically showing that the chosen model does not, in fact, correspond to the Maxwell--Cattaneo hyperbolic (wave) theory of heat transfer.
\end{abstract}

% PACS are no longer used in Phys. Rev., see https://journals.aps.org/PACS
%\pacs{} 
                  
% Keywords are not printed in Phys. Rev. articles                                    
%\keywords{} %Use showkeys class option if keyword display desired

\maketitle

\paragraph*{Introduction.}
This Comment presents a critique of the recent {\it Physical Review Applied} publication \cite{Farhat2019}, the focus of which is a proposed cloaking scheme for thermal waves in rigid solids.  To begin, it is instructive to briefly review the topic of hyperbolic heat transport, i.e., the theory of \emph{heat waves} \cite{JP89a,*JP89b,S11}, in rigid solids.

Consider a thermally conducting, homogeneous and isotropic, rigid solid at rest. As first suggested by theory, and subsequently confirmed by experiment, at sufficiently low temperatures the transport of heat in such bodies occurs \emph{not} via diffusion, the mechanism underlying Fourier's law for the thermal flux, but instead by the propagation of thermal waves (or \emph{second sound}) \cite{JP89a}. Many constitutive relations have been proposed to describe this phenomenon \cite{S11}. Perhaps the best known is the Maxwell--Cattaneo (MC) law~\cite{M67,*C48}, which in the present context reads
\begin{equation}\label{eq:MC_law}
	\left(1+\tau_{0}\frac{\partial }{\partial t}\right)\bm{\Phi} =-\kappa_{0} \bm{\nabla} T.
\end{equation}
Unsurprisingly, the history of this relation is complex: there exists Russian-language literature describing a similar flux law prior to Cattaneo (but, of course, after Maxwell) \cite{Sobolev,*B03}. Here, $T=T(\bm{x},t)$  and $\bm{\Phi}=\bm{\Phi}(\bm{x},t)$ denote the absolute temperature and the thermal flux vector, respectively, where  $\bm{x} = (x,y,z) \in \mathbb{R}^3$.  As in \cite{Farhat2019},  $\tau_{0}(>0)$ is the thermal relaxation time for phonon processes that do not conserve phonon momentum~\cite{JP89a}, and $\kappa_{0}(>0)$ is the thermal conductivity of the solid under consideration.
 
Equation~\eqref{eq:MC_law}, which reduces to Fourier's law on setting $\tau_{0} \equiv 0$, is the latter's simplest generalization that yields a hyperbolic thermal transport equation, unlike the parabolic transport equation that stems from Fourier's law. Therefore, the MC law overcomes the so-called ``paradox of diffusion''---the philosophically problematic implication that thermal disturbances in continuous media propagate with infinite speed under Fourier's law.

Delving into the critique of the study carried out in \cite{Farhat2019} will further illustrate these notions. Unless otherwise stated the same notation used in \cite{Farhat2019} is employed herein. First, observe that in \cite[Eqs.~(3)]{Farhat2019}, the energy balance equation is incorrectly stated; specifically, its source term, which is denoted here by $\mathcal{S}=\mathcal{S}(\bm{x},t)$, is missing and  the $\partial T/\partial t$ term it contains should be multiplied by the product $\varrho c_{\rm p}$ to ensure dimensional consistency. Second, but more troubling, the ``flux diffusion'' term, which is introduced into the MC law~\eqref{eq:MC_law}, appears to be an attempt to introduce some of the features of the flux relation of Guyer and Krumhansl (GK) \cite[Eq.~(59)]{GK66b}, which was  derived by GK from the linear Boltzmann equation. Note that under the GK model, heat flow is \emph{not} necessarily down the temperature gradient and certainly ``will not permit the propagation of [heat] waves'' \cite[p.~46]{JP89a} unless the ``flux diffusion'' terms are neglected.

Now, with the required corrections made to yield the actual GK flux law, \cite[Eqs.~(3)]{Farhat2019} can be expressed as
\begin{subequations}\label{Sys-1}
\begin{align}\label{Sys-1:ConvEgy_1}
\varrho c_{\rm p} \frac{\partial T}{\partial t} &=-\bm{\nabla \cdot \Phi} + \mathcal{S},\\
\label{eq:GK_law}
\underbrace{\left[1+\tau_{0}\frac{\partial }{\partial t} -\tau_{0}\sigma_{0}({\Delta} + 2\bm{\nabla\nabla\cdot}) \right]}_{=: \mathscr{K}} \bm{\Phi} &=-\kappa_{0} \bm{\nabla} T,
\end{align}\end{subequations}
where $\Delta := \bm{\nabla\cdot\nabla}$ is the Laplacian operator. In Eqs.~\eqref{Sys-1},  $c_{\rm p}$ (not $c_{\rm v}$, see \cite[p.~9]{cj59}) and $\varrho$  are the specific heat at constant pressure and the mass density, respectively, of the solid under consideration. Next, it is easily established from \cite[Sect.~IV]{JP89a} that $\sigma_{0}=(1/5)\tau_{N}V^2$, where $\tau_{N}$ is the relaxation time for $N$-processes and $V$ carries SI units of m s$^{-1}$. Therefore, the SI units of $\sigma_0$ are m$^2$ s$^{-1}$; \emph{not} W m$^{-1}$ K$^{-1}$, as reported in \cite[p.~3]{Farhat2019}. The operator $\mathscr{K}$ here differs from its counterpart in \cite{Farhat2019} in that the latter is missing $\bm{\nabla \nabla\cdot}$.

\begin{remark}
Again, while Eq.~\eqref{eq:MC_law} is the $\sigma_{0} \equiv 0$ special case of  Eq.~\eqref{eq:GK_law}, it is important to stress that it is \emph{incorrect} to regard the latter as exhibiting a small, ``innocent'' correction to the former. As shown below, the MC flux law~\eqref{eq:MC_law} predicts heat \emph{waves} (hyperbolic thermal transport equation), while the GK flux law~\eqref{eq:GK_law} predicts heat \emph{diffusion} (parabolic thermal transport equation).
\end{remark}

\begin{remark}
Observe that, as a result of erroneously dropping $\varrho c_\mathrm{p}$ in the energy balance, many equations in \cite{Farhat2019} are dimensionally inconsistent and, therefore, devoid of physical meaning. For example, consider \cite[Eq.~(4)]{Farhat2019}. The first and second terms on the left-hand side (LHS) have units K s$^{-1}$, while the third and fourth terms on the LHS have units W m$^{-3}$.
\end{remark}

\paragraph*{The thermal transport equation.} 
As in \cite{Farhat2019}, regard all coefficients as constant and proceed to  eliminate $\bm{\Phi}$ between the equations of Eqs.~\eqref{Sys-1}, assuming sufficient smoothness of the dependent variables.  The first step in this process is employing Eq.~\eqref{Sys-1:ConvEgy_1} to recast Eq.~\eqref{eq:GK_law} as
\begin{multline}
\label{eq:GK_law_K}
\underbrace{\left[1+\tau_{0}\frac{\partial }{\partial t} -\tau_{0}\sigma_{0}{\Delta}\right]}_{=: \mathscr{H}_{\sigma_{0}}} \bm{\Phi} 
=-\kappa_{0} \bm{\nabla} T \\
+ 2\tau_{0}\sigma_{0}\left[\bm{\nabla}\mathcal{S} -  \varrho c_{\rm p}\frac{\partial (\bm{\nabla} T)}{\partial t}\right].
\end{multline}
Next, after applying $\mathscr{H}_{\sigma_{0}}$ to Eq.~\eqref{Sys-1:ConvEgy_1}, and then using Eq.~\eqref{eq:GK_law_K}, one obtains the thermal transport equation 
\begin{equation}\label{eq:GK_Eq3D}
\frac{\partial T}{\partial t}+\tau_{0}\frac{\partial^2 T}{\partial t^2} = \tau_{0}\tilde{\sigma}_{0} \frac{\partial (\Delta T)}{\partial t} + \varkappa_{0} \Delta T 
+ \frac{1}{\varrho c_{\rm p}} \mathscr{H}_{\tilde{\sigma}_{0}}[\mathcal{S}],
\end{equation}
where $\varkappa_{0} := \kappa_{0}/(\varrho c_{\rm p})$ is the thermal \emph{diffusivity}  and, for convenience, $\tilde{\sigma}_0 := 3\sigma_0$ has been defined. As the right-hand side (RHS) of \cite[Eq.~(4)]{Farhat2019} is not acted upon by the operator $\mathscr{H}_{\tilde{\sigma}_{0}}$, nor multiplied by  $1/(\varrho c_{\rm p})$, and its LHS contains $\sigma_{0}$, not $\tilde{\sigma}_{0}$,  Eq.~\eqref{eq:GK_Eq3D} above is the \emph{corrected} version of \cite[Eq.~(4)]{Farhat2019}.

\begin{remark}
The $\tilde{\sigma}_0 \equiv 0$, source-free version of Eq.~\eqref{eq:GK_Eq3D} is  the  multidimensional version of the \emph{damped wave equation} \cite[p.~42]{JP89a}, which predicts that thermal signals (disturbances) propagate at a \emph{finite} characteristic speed of $c_{0} := \sqrt{\varkappa_{0}/\tau_{0}}$ (see also \cite{BH69,*BH71}). Meanwhile, the $\tilde{\sigma}_0 > 0$, source-free version of Eq.~\eqref{eq:GK_Eq3D} is a multidimensional \emph{Jeffreys-type equation}, which predicts an \emph{infinite} speed of propagation of signals \cite[p.~46]{JP89a}.
\end{remark}

To demonstrate this important difference between wave-like and diffusive  thermal transport (see also \cite{ETS_ICC}), but in a slightly simpler way, consider a related one-dimensional (1D) initial-boundary value problem~(IBVP) posed by Tanner~\cite{Tanner62} for the Jeffreys-type equation arising in the context of viscoelasticity. (This IBVP is also the one considered in \cite{BH69,*BH71} for the damped wave equation of hyperbolic heat conduction.) Recasting Tanner's problem in the present notation:
\begin{subequations}\label{IBVP:Tanner}
\begin{align}
\label{IBVP}
& \frac{\partial T}{\partial t}+\tau_{0}\frac{\partial^2 T}{\partial t^2} = \tau_{0} \tilde{\sigma}_0 \frac{\partial^3 T}{\partial t\partial x^2} + \varkappa_{0}\frac{\partial^2 T}{\partial x^2}, \quad (x,t)\in \Omega,\\
& T(0,t) = T_{0}H(t),\quad \lim_{x \to \infty}T(x, t) = 0, \quad  t >0,\label{BC}\\
& T(x,0) =  \frac{\partial T}{\partial t}(x,0) = 0, \quad  x >0,\label{IC}
\end{align}
\end{subequations}
where $H(\cdot)$ denotes the Heaviside unit step function, $\Omega := (0, \infty)\times (0, \infty)$ is the space-time domain of interest, and the constant $T_{0}(>0)$ is the amplitude of the inserted thermal signal. Following Tanner~\cite{Tanner62}, one can apply the Laplace transform to Eq.~\eqref{IBVP} and its boundary conditions (BCs) \eqref{BC}. This IBVP correspond to a \emph{heat pulse} experiment \cite{Van2017,Berezovski}. After making use of the initial conditions \eqref{IC}, and then  solving the resulting subsidiary equation subject to the (transformed) BCs, one obtains an algebraic expression that can inverted back to the time domain. This exact inverse is known \cite{Tanner62}, with several other representations summarized in \cite{C10}.

\begin{figure}
\includegraphics[width=0.9\columnwidth]{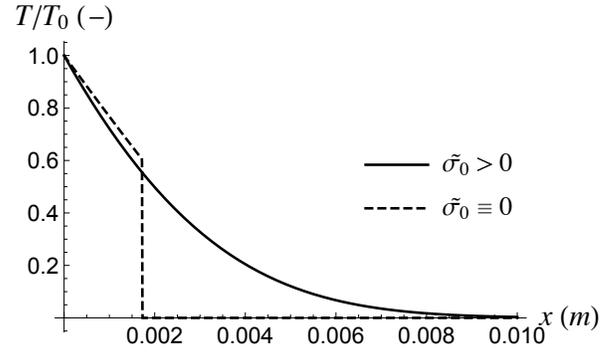}
\caption{Wave-like (dashed curve, $\tilde{\sigma}_0 \equiv 0$) versus diffusive (solid curve, $\tilde{\sigma}_0 = 6.4015\times10^{-6}$ m$^{2}$ s$^{-1}$) spread of heat from the sudden imposition of a temperature jump at $(x,t)=(0,0)$. Here, $\tau_0=0.991$ s and $\varkappa_0 = 2.95 \times 10^{-6}$ m$^2$ s$^{-1}$, as given in \cite{Van2017} for limestone. This plot is at $t=1$ s, i.e., after approximately a single thermal relaxation time. ~\label{fig:example}}
\end{figure}

To illustrate this fundamental difference between the thermal transport described by the MC law and the GK-type law used in \cite{Farhat2019}, the respective \emph{exact} solutions of IBVP~\eqref{IBVP:Tanner} are shown in Fig.~\ref{fig:example} for $\tilde{\sigma}_0\equiv 0$ and $\tilde{\sigma}_0 > 0$, respectively. The integral expression for the exact solution obtained by Tanner~\cite{Tanner62} is evaluated numerically using {\sc Mathematica}'s {\tt NIntegrate} subroutine, to arbitrary precision, on a finite grid of $x$ values \cite{C10}. Figure~\ref{fig:example} shows that under the MC law (dashed curve), the heat pulse has only propagated slightly less than 2 mm into the domain. Meanwhile, under the GK law (solid curve), the normalized temperature $T/T_0$ is non-zero \emph{everywhere} in the domain. Since the physical parameter values given in \cite{Farhat2019} are incorrect/inconsistent, to generate the plots in Fig.~\ref{fig:example}, the values for limestone are taken from \cite{Van2017}, wherein it was experimentally demonstrated that material micro-structure can lead to GK-type heat conduction at room temperature (see also \cite[Ch.~9]{Berezovski}).

To summarize: the behavior of the  $\tilde{\sigma}_0 >  0$ case is seen to be strictly diffusive (similar to the solution of the 1D thermal diffusion equation arising from Fourier's law); the signal applied at $x=0$ is ``felt'' \emph{instantly}, but equally, at every point in the half-space $x>0$. In contrast, the $\tilde{\sigma}_0 \equiv  0$ case predicts a thermal shock-front, of magnitude $T_{0} \exp[-t/(2\tau_{0})]$, propagating (to the right) with \emph{finite} speed $c_{0} \approx 1.74 \times 10^{-3}$ m s$^{-1}$ (for the chosen parameter values); see also \cite[p.~545]{BH69} and \cite[p.~45]{JP89a}.

\paragraph*{Harmonic disturbances.} 
Returning to Farhat \textit{et al.}'s analysis, set $\mathcal{S}(\bm{x},t) = 0$ and assume $T(\bm{x},t)=\Theta(\bm{x})\exp(-\ri\omega t)$, where $\omega (>0)$ is the angular frequency of some thermal disturbance impacting the solid in question. Under these assumptions, Eq.~\eqref{eq:GK_Eq3D} is reduced to the (source-free) Helmholtz equation 
\begin{equation}\label{eq:Helmholtz_3D}
\Delta \Theta +\left(\frac{\tau_{0}\omega^2+i\omega}{\varkappa_{0} - \ri\tau_{0}\tilde{\sigma}_{0}\omega } \right) \Theta = 0.
\end{equation}
It should be noted that, in \cite[Eq.~(5)]{Farhat2019}, ``$T$'' is reused instead of introducing a new (time-independent) function such as $\Theta$ herein.  \cite[Eq.~(5)]{Farhat2019} also incorrectly features the thermal conductivity, with its subscript (``0'') missing, in place of the thermal \emph{diffusivity} $\varkappa_0$.

Consider plane wave propagation in a direction set by the unit vector $\hat{\bm{u}}$. Then, on setting  $\Theta(\bm{x}) = \Theta_0 \exp(\ri k_{0}\hat{\bm{u}}\cdot\bm{x})$, Eq.~\eqref{eq:Helmholtz_3D} yields the dispersion relation
\begin{equation}\label{eq:Helmholtz_3D_disp_sq}
k_{0}^2(\omega) = \underbrace{\frac{\tau_{0}\omega^2(\varkappa_{0}-\sigma_{0})}{\varkappa_{0}^2+\tau_{0}^2\tilde{\sigma}_{0}^2\omega^2}}_{ =:\mathfrak{a}} + \ri \underbrace{\frac{\omega (\varkappa_{0}+\tau_{0}^2\tilde{\sigma}_{0}\omega^2)}{\varkappa_{0}^2+\tau_{0}^2\tilde{\sigma}_{0}^2\omega^2}}_{=:\mathfrak{b}} \, ,
\end{equation}
where, $\ri=\sqrt{-1}$, $\Theta_0 > 0$ and $k_{0}\in\mathbb{C}$. Enforcing $\Theta < \infty$ as $|\bm{x}|\to\infty$ (and, also, since $\mathfrak{b}>0$) requires $\Imag(k_0)\geq 0$, then it is readily established that
\begin{equation}\label{eq:Helmholtz_3D_disp}
k_{0}(\omega) = \sqrt{\frac{\mathfrak{a}+\sqrt{\mathfrak{a}^2+\mathfrak{b}^2}}{2}} + \ri \sqrt{\frac{-\mathfrak{a}+\sqrt{\mathfrak{a}^2+\mathfrak{b}^2}}{2}}.
\end{equation}
 
\begin{figure}
\includegraphics[width=0.9\columnwidth]{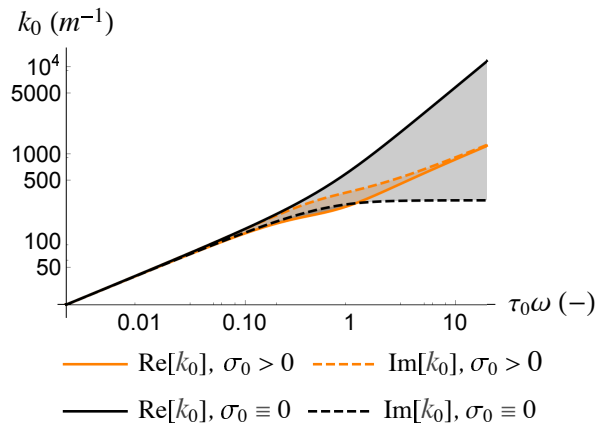}
\caption{Real and imaginary parts of the wavenumber $k_0$ (dark colors for the MC law and light colors for the GK law) as computed from Eq.~\eqref{eq:Helmholtz_3D_disp} for the same parameter values used to generate Fig.~\ref{fig:example}.~\label{fig:dispersion}}
\end{figure}

The dispersion relation in Eq.~\eqref{eq:Helmholtz_3D_disp}, as well as the $\tilde{\sigma}_0 \equiv 0$ reduction to its version under the MC law, are illustrated in Fig.~\ref{fig:dispersion}. Although the dark contours may look similar to \cite[Fig.~1(b,bottom)]{Farhat2019}, observe that $\Real(k_0)<\Imag(k_0)$ for the chosen set of (realistic) physical parameters, contrary to what is shown in \cite{Farhat2019}. However, $\Real(k_0)>\Imag(k_0)$ does hold true under the MC law ($\tilde{\sigma}_0 \equiv 0$). Furthermore, in this case of hyperbolic heat (wave) transfer, the scattering and absorption are \emph{not} balanced, because $\Imag(k_0)\to(4\varkappa_0\tau_0)^{-1/2} = const.$ but $\Real(k_0)\sim \tau_0\omega/\sqrt{\varkappa_0\tau_0}$, as $\tau_0\omega\to\infty$.

\paragraph*{Other issues.}
In addition to those detailed above, the following other errors/issues were noticed in \cite{Farhat2019}:
\begin{itemize}
\item[(i)] In \cite[p.~2]{Farhat2019}, it is claimed that the term proportional to $\sigma_{0}$ is necessary to ``make the discretizing process asymptotically stable.'' Leaving aside the unclear meaning of ``asymptotically'' in this context, this statement is false. There is no difficulty whatsoever in discretizing a hyperbolic heat transport equation by any number of methods, as has been known for over three decades (see, e.g., \cite{Carey1982,*Glass1985}, but note that modern schemes \cite{LeVeque} should be used nowadays).

\item[(ii)] Below \cite[Eq.~(5)]{Farhat2019}, it is stated that ``$k_0$ is a
complex number for all frequencies [under the GK-type flux law], which is markedly different from classical heat waves (Fourier transfer).'' Setting aside the fact that heat waves are \emph{impossible} under Fourier's law, it is clearly seen, on setting $\tau_0 \equiv 0$ in Eq.~\eqref{eq:Helmholtz_3D_disp_sq}, that $k_{0}(\omega) = \sqrt{\ri\omega/\varkappa_0} = (1+\ri)\sqrt{\omega/(2\varkappa_0)}\in\mathbb{C}$; i.e., there is no marked difference between Fourier and non-Fourier heat flux laws in this regard.

\item[(iii)] The unknown coefficients in the expansions in \cite[Eqs.~(8) and (9)]{Farhat2019} are found by applying a boundary condition involving  ``the temperature field $T$, as well as
its flux $\kappa\bm{\nabla}T$.''  Under the MC law, the heat flux is \emph{not} (with misprints corrected) $-\kappa_{0}\bm{\nabla}T$, as it would be under \emph{Fourier}'s law; rather, it is the expression obtained by solving Eq.~\eqref{eq:MC_law} for $\bm{\Phi}$.  In the case of harmonic time-dependence, for which $\bm{\Phi}(\bm{x},t)=\bm{F}(\bm{x})\exp(-\ri\omega t)$, specifying the flux at the boundary of some spatial domain $\mathcal{D}\subset \mathbb{R}^{3}$, under the MC law,  would correspond to specifying 
\begin{equation}\label{eq:MC_flux BC}
\bm{F} = -\left(\frac{1-\ri\omega \tau_{0}}{1+\omega^2 \tau_{0}^{2}}\right) \kappa_{0} \nabla \Theta \quad \text{on}\quad \bm{x} \in \partial \mathcal{D}.
\end{equation}
The corresponding expression under the GK flux law~\eqref{eq:GK_law} is lengthier. This error in imposing the BCs on the series expansion puts into question all subsequent results in \cite[Sec.~III and IV]{Farhat2019}.

\item[(iv)] The conclusion of \cite[p.~7]{Farhat2019} states that ``the Fourier heat equation is not frame invariant.'' However, this statement is false. The thermal transport equation under Fourier's law is indeed frame-invariant because the material derivative $DT/Dt := \partial T/\partial t + \bm{v}\bm{\cdot}\bm{\nabla}T$ is featured on the LHS of Eq.~\eqref{Sys-1:ConvEgy_1} in its derivation for heat transfer in a moving  (or deforming) medium with velocity $\bm{v}$ (see, e.g., \cite[Sec.~7.1]{Jog}), whence $D T/Dt \equiv \partial T/\partial t$ if $\bm{v} = \bm{0}$ (stationary conductor). Furthermore, the frame-indifferent formulation of the MC law is misattributed in \cite{Farhat2019}; it was, in fact, derived in \cite[Ref.~35]{Farhat2019} not \cite[Ref.~43]{Farhat2019}.

\item[(v)] The word ``photon(s)'' should be replaced with ``phonon(s)'' everywhere in \cite{Farhat2019}, given that the context is heat conduction, not electromagnetism.
\end{itemize}

\paragraph*{Conclusion.}
On the basis of the above-identified errors and stated criticisms, it must be concluded that Farhat \textit{et al.}~\cite{Farhat2019} have failed to provide ``the first demonstration of scattering cancellation cloaking for heat \emph{waves} [emphasis added] obeying the Maxwell--Cattaneo transfer (sic) law.''  It would, therefore, be of interest to re-do the study attempted in \cite{Farhat2019}, with the correct physical model [i.e., the $\sigma_{0} \equiv 0$ special case of Eqs.~\eqref{Sys-1}], the correct boundary conditions and correct parameter values, to determine whether cloaking is possible (or not).

Finally, with regards to item (iv) above, it is appropriate to mention the promulgation of dubious results. The reworking of classical results under the frame-indifferent generalization of the MC law has generated a large and scientifically/mathematically questionable literature \cite{p17,*p19,*p19b}, which it is easily verified that \cite[Ref.~39]{Farhat2019} is related to.

\begin{acknowledgements}
Contributions to an earlier draft by an anonymous colleague are acknowledged.
\end{acknowledgements}

\bibliography{PRAppl_Comment.bib}

\end{document}